# Counting chickens before they hatch: reciprocal consistency of calibration points for estimating divergence dates

David A. Morrison

Section for Parasitology (SWEPAR)
Department of Biomedical Sciences and Veterinary Public Health
Swedish University of Agricultural Sciences, 751 89 Uppsala, Sweden
Email: David.Morrison@bvf.slu.se

**Abstract**

There has been concern in the literature about the methodology of using secondary calibration timepoints when estimating evolutionary divergence dates. Such timepoints are divergence time estimates that have been derived from one molecular data set on the basis of a primary external calibration timepoint, and which are then used independently on a second data set. Logically, the primary and secondary calibration points must be mutually consistent, in the sense that it must be possible to predict each time point from the other. However, the attempt by Shaul and Graur (2002, Gene 300: 59–61) to assess the reliability of secondary timepoints is flawed because they presented time estimates without presenting confidence intervals on those estimates, and so it was not possible to make any explicit hypothesis tests of divergence times. Also, they inappropriately excluded some of the data, which leads to a very biased estimate of one of the divergence times. Here, I present a re-analysis of the same data set, with more appropriate methodology, and come to the conclusion that no inconsistencies are involved. However, it is clear from the analysis that molecular data often have such large confidence intervals that they are uninformative, and thus cannot be used for reliable hypothesis tests.

## 1. Introduction

In a recent methodological exchange concerning the use of molecular data to estimate times of evolutionary divergence between taxa [3, 5, 8], the paper by Shaul and Graur [17] was cited with reference to the use of secondary calibration timepoints. Secondary (or indirect) calibration points are divergence-time estimates that have been derived from a primary calibration point, which was in turn derived from some independent source of historical dates (e.g. palaeontological or biogeographic evidence). The secondary calibration point is then used in molecular data sets where use of the primary calibration point is inappropriate (e.g. one or more of the taxa involved in the primary calibration point do not appear in the data set).

The issue raised by Shaul & Graur is that the secondary calibration time is not independent of the primary calibration time (since it is derived from it), and thus it needs to be mutually consistent with the primary point. That is, use of the secondary calibration point in a calculation must be capable of leading accurately back to the primary calibration point. For example, if we were to use a primary calibration time of (say) 310 million years ago (MYA) for the bird–mammal divergence to derive a secondary calibration time of (say) 110 million years ago for the primate-rodent divergence, then subsequent use of this primate-rodent divergence time as a calibration point must predict a bird–mammal divergence time of 310 MYA. If this does not happen then the two calibration points are not consistent, and the secondary calibration point will be of little practical value as a substitute for the primary point.

This is clearly an important methodological issue, and so Shaul & Graur attempted to substantiate their point by re-analysing the data of Wang et al. [21], and concluded that the secondary calibration time used throughout the latter study was inconsistent with the primary calibration time. However, here I point out that the methodology used by Shaul & Graur was unsuitable as a test of their idea. In particular, they presented time estimates without presenting confidence intervals on those estimates, and so it is not possible to make any explicit hypothesis tests of divergence times. Also, they inappropriately excluded some of the data, which leads to a very biased estimate of one of the divergence times. Here, I present a re-analysis of the same data set that they used, with more appropriate methodology, and I come to quite different conclusions.

## 2. Data and methods

The data analysis essentially repeats that of Shaul & Graur, based on the data described by Wang et al. [21]. The data consist of amino acid sequences for each of 29 homologous proteins for each of three vertebrate taxa: a bird (chicken), a primate (human) and a rodent (mouse or rat). The objective of the analysis is to derive an estimate of the secondary calibration time (primate–rodent divergence) based on the primary calibration time (bird–mammal divergence) and, reciprocally, to estimate the primary calibration time based on the secondary calibration time. The calibration times used are 310 MYA for the bird–mammal divergence (T1) and 110 MYA for the primate–rodent divergence (T2).

I calculated the sequence divergence between the pairs of taxa as the poisson-corrected distance and its variance as described by Nei et al. [11], using the MEGA ver. 2.1 program of Kumar et al. [9]. The two divergence times based on the data for each gene were then calculated using the formulae presented by Shaul & Graur. The standard error of each time estimate was calculated by combining the errors for the component genetic distances, using standard methods based on quadrature [19]. These standard errors were then used to calculate confidence intervals for each estimate; this is the first significant point of departure from the methods employed by Shaul & Graur. Note that the calibration time used in each calculation is assumed to be without error (i.e. it is a constant and does not contribute to the standard error of the final estimated time), which is an appropriate assumption for the tests performed here but will not be appropriate in general for time estimates [5].

The second point of departure from the methods employed by Shaul & Graur is that I have used the geometric mean and its confidence interval throughout, rather than the arithmetic mean and confidence interval. There has been much concern in the literature about the fact that the frequency distribution of divergence-time estimates is not symmetrical [8, 11, 14], and various strategies for dealing with this have been proposed. Here, I have followed the argument of Morrison [10] that this frequency distribution is basically lognormal and that therefore the geometric mean is the most appropriate measure of central location. It is straightforward to convert an arithmetic mean and standard error onto a logarithmic scale, and thus to derive the geometric mean and its confidence interval. However, for these data it makes very little difference whether an arithmetic or geometric mean is used, and it does not affect the overall conclusions.

Shaul & Graur devised a consistency test for the primary and secondary calibration times consisting of two parts, which could be applied to the data for each gene: (1) T1 < T2 (i.e. the primate–rodent divergence pre-dates the bird–mammal divergence); and (2) T1 implies that T2 ≈ 310 MYA (i.e. use of the secondary calibration time yields an estimate of the bird–mammal divergence that is similar to the primary calibration time).

## 3. Results and Discussion

### 3.1 Consistency Test Part (1)

The essential methodological limitation of the analysis performed by Shaul & Graur is that the time estimates they produced for T1 and T2 were point estimates with no associated estimation of statistical confidence. Thus, there was no objective basis for comparing the two times T1 and T2, as required by part (1) of their consistency test. This led them, for example, to reject the hypothesis that T1<T2 for the tryptophan hydroxylase gene data (186>184 MYA) but to accept it for the ferritin gene data (181<188 MYA) when these proteins obviously produce very similar estimates, which are well within the expected inaccuracy of the time-estimation method.

A revised version of the 29 pairs of time estimates are therefore shown in Table 1a, along with a 95% confidence interval for each estimate. Also shown are the results of appropriate statistical hypothesis tests based on these confidence intervals — if a pair of 95% confidence intervals do not overlap then we can reject the hypothesis that the two estimates are equal at $P=0.05$. On this basis, T2>T1 for 14 of the 29 gene sequences and in no case is T1>T2. Thus, these data pass part (1) of the consistency test. This conclusion contrasts with that of Shaul & Graur, who decided that 7 of the 29 genes failed the test. This difference in the conclusions is not a result of my having used geometric rather than arithmetic means, as use of the latter procedure leads to the same result (Table 1a), the only difference being that use of the arithmetic means and confidence intervals leads to a somewhat less powerful test (i.e. there are 12 significant results instead of 14). Thus, test part (1) produces robust results.

### 3.2 Consistency Test Part (2)

In order to examine part (2) of their consistency test, Shaul & Graur deleted the data for the 7 genes that failed part (1) of the test and then deleted a value (for Na-K ATPase beta) detected as an outlier (using Grubb's outlier test) compared to the remaining T2 values. From the remaining 21 genes they calculated an average T2 value and its 95% confidence interval.

However, there is clearly no basis for deleting the group of 7 genes, irrespective of whether they failed part (1) of the test or not. The data for the 29 genes represent a sample taken from a statistical population and thus can be expected to show variability due to stochastic variation around the population mean — deleting the 7 smallest values (which is by definition what these 7 time estimates are) must produce a biased estimate of that mean. Furthermore, Grubb's test is based on comparing each value to an expected normal (i.e. gaussian) probability distribution; and deleting the smallest values will create an even more skewed version of what is already expected to be a skewed frequency distribution (i.e. lognormal), which will invalidate the use of Grubb's test (unless a logarithmic data transformation is used).

A revised pair of confidence intervals is therefore shown in Table 1a, incorporating the time estimates from all of the proteins. The interval for T2 firmly includes the predicted T2=310 MYA, and the interval for T1 also includes the predicted T1=110 MYA. This conclusion contrasts with that of Shaul & Graur, whose biased interval for T2 was 315–471 MYA. Note that this latter interval should properly be compared to the interval shown in Table 1a without the outlier — the time estimate from the Na-K ATPase beta chain is also an outlier when compared to the lognormal probability distribution fitted to the data of Table 1a. In this case, there is even less overlap with the interval produced by Shaul & Graur. However, the confidence intervals produced without the outlier still include the predicted values. Thus, test part (2) also produces robust results.

I thus conclude that the data of Wang et al. [21] do not provide a suitable example of the potential inconsistency of secondary calibration points, at least as analysed using the methodology of Shaul & Graur. In this regard, Hedges et al. [7] reported that 118 of their 120 proteins showed consistency (i.e. T2>T1). Unfortunately, no statistical evidence (such as confidence intervals) was used to substantiate this claim.

## *3.3 Problems with Secondary Calibration*

However, my conclusion does not mean that there are no problems with the use of secondary calibration times. I think that it is particularly important to note that 15 of the 29 protein data sets analysed here did not contain sufficient evolutionary information to be able to reject the hypothesis that the primate–rodent divergence post-dates the bird–mammal divergence (Table 1a), in spite of the fact that 200 million years are usually presumed to have occurred between these two events. So, even though the data are not inconsistent, they are not particularly informative either, because the confidence intervals are often too large for reliable hypothesis tests. This point is particularly stressed by Graur and Martin [5].

Furthermore, 12 of the 29 data sets produced confidence intervals that allow us to reject the hypothesis that T1=110 MYA, and 12 of the data sets (not always the same ones) produced confidence intervals that allow us to reject the hypothesis that T2=310 MYA (Table 1a), even though the overall averages across the proteins are consistent with these two dates. This emphasizes the need for multi-gene data sets if reliable time estimates are to be forthcoming. However, even using all 29 proteins the confidence interval for the bird–mammal divergence is very large (238–384 MYA). This inexactness is consistent with the results of Glazko et al. [3], who obtained estimates (without confidence intervals) of 292 and 329 MYA based on two other calibration times. Clearly, the time of the bird–mammal divergence is still uncertain.

As another cautionary point, it is worth noting just how labile the time estimates can be as a result of the particular gene sequences chosen for analysis. Several of the proteins used here have isoforms (e.g. expressed in different tissues coded by different loci), and only one of these forms was included in the data of Wang et al. [21]. Table 1b shows the time estimates produced by three such protein sequences. In all three cases the results vary among isoforms as far as test part (1) is concerned, and thus the results of this test would differ depending on which isoform is chosen for inclusion. Furthermore, the outlying value for test part (2) results from choosing the form of Na-K ATPase beta with the most extreme time estimates. However, none of the Na-K ATPase beta isoforms produced results that are compatible with the predicted values of T1 and T2.

*3.4 Methodology for Confidence Intervals*

As far as methodology is concerned, time estimation using the standard procedures clearly involves combining variables that each have associated measurement errors, although this point has largely been ignored by practitioners (an argument made strongly by [5]). The combining of these errors needs to be done using some objective method, rather than the apparently additive manner used by Graur and Martin [5]. The method used here (i.e. quadrature) assumes that the covariances between the genetic distances are zero, so that they can thus be disregarded when calculating the overall error of the time estimate. This assumption may be unrealistic, due to the non-independence of evolutionary relationships (i.e. taxa are related by a tree and therefore the distances are partly shared along the branches). This means that the standard errors are likely to be mis-estimated, and the exact formula of Goodman [4] might to be more appropriate. However, the covariances cannot be estimated accurately unless measurements come in pairs, which means that exactly the same taxa and genes need to be used for all measurements. Alternatively, more sophisticated methods from the general field of error analysis and error propagation, involving the sensitivity coefficients for each component, could be used [2, 12]. However, these methods still assume that the measured quantities are sampled from a normal distribution, which is unlikely to be true of genetic distances (e.g. they are more likely to follow a gamma distribution, which will only approximate a normal distribution under certain circumstances). The effect of this on the final time estimates is not clear.

In fact, it might be better to try to put confidence intervals directly on the time estimates themselves, rather than combining the component errors. Several tree-based methods have recently been developed to do this [1, 11, 16, 20] especially with reference to multi-locus data sets. However, these methods all assume that the phylogenetic tree is fixed (i.e. known without error). This is also an unrealistic assumption, whether it refers only to the topology or to the branch lengths as well; and so this might not be a practical gain at all, as far as accurate error estimation is concerned.

The absolute importance of providing confidence intervals for evolutionary time estimates has been a major issue raised by a number of workers over nearly a decade [5, 6, 13, 15, 18], and I wish to re-emphasize it here — no meaningful comparison of evolutionary times is possible in the absence of confidence intervals on the time estimates. Indeed, rather surprisingly Shaul & Graur make this the very final point of their paper, apparently failing to note its inconsistency with their own analysis.

Table 1. Estimates and statistical tests of two divergence times.

| Locus name | Geometric mean and 95% confidence interval | | Statistical significance [b] | | | |
| --- | --- | --- | --- | --- | --- | --- |
| | | | Lognormal [c] | | Normal [d] | |
| | T1 (MYA) [a] | T2 (MYA) [a] | T2>T1 | T1>T2 | T2>T1 | T1>T2 |

*(a) Data of Wang, Kumar and Hedges (1999)*

| Locus name | T1 (MYA) | T2 (MYA) | T2>T1 (LN) | T1>T2 (LN) | T2>T1 (N) | T1>T2 (N) |
| --- | --- | --- | --- | --- | --- | --- |
| Aldehyde dehydrogenase | 143–215–325 | 100–151–229 | | | | |
| Aldolase | 28–62–138 | 209–467–1043 | * | | | |
| Alkaline phosphatase | 71–103–150 | 219–318–462 | * | | * | |
| Alpha actinin | 138–261–495 | 62–117–222 | | | | |
| Amidophosphoribosyl transferase | 66–104–164 | 197–310–490 | * | | * | |
| Aminolevulinate synthetase | 146–204–286 | 116–162–227 | | | | |
| Aspartate aminotransferase | 85–133–209 | 155–243–381 | | | | |
| Dihydrofolate reductase | 62–112–202 | 154–279–505 | | | | |
| Disulfide isomerase | 71–112–175 | 185–289–453 | * | | | |
| DNA polymerase gamma | 92–131–185 | 183–263–378 | | | | |
| Enolase | 114–213–398 | 77–145–270 | | | | |
| Ferritin heavy chain | 65–162–405 | 68–169–423 | | | | |
| Fructose-2,6-bisphosphatase | 39–63–105 | 305–503–830 | * | | * | |
| Furin | 55–80–116 | 284–411–596 | * | | * | |
| Glutamate dehydrogenase | 16–38–92 | 319–739–1712 | * | | * | |
| Glutamine synthetase | 105–181–312 | 101–175–301 | | | | |
| Glyceraldehyde-3-phosphate dehydrogenase | 107–214–430 | 70–140–282 | | | | |
| Lactate dehydrogenase | 63–114–206 | 151–273–494 | | | | |
| Na-K ATPase alpha chain | 60–118–235 | 129–255–505 | | | | |
| Na-K ATPase beta chain | 7–14–27 | 1106–2190–4337 | * | | * | |
| P53 | 72–102–145 | 228–324–461 | * | | * | |
| P65 | 39–52–72 | 463–639–881 | * | | * | |
| Phosphoenolpyruvate carboxykinase | 114–166–243 | 135–197–288 | | | | |
| Phosphoglycerate kinase | 26–53–109 | 283–567–1135 | * | | * | |
| Pyruvate kinase | 40–69–119 | 264–458–794 | * | | * | |
| Transcription factor Eryf1 | 32–51–79 | 408–639–1001 | * | | * | |
| Transglutaminase | 89–113–144 | 234–297–378 | * | | * | |
| Triosephosphate isomerase | 59–124–258 | 117–247–524 | | | | |
| Tryptophan hydroxylase | 118–182–282 | 115–178–275 | | | | |
| Sum | | | 14 | 0 | 12 | 0 |
| Geometric average and confidence interval | 82–104–133 | 238–302–384 | | | | |
| — minus outlier | 92–112–137 | 231–282–343 | | | | |

Table 1. continued

| Locus name | Geometric mean and 95% confidence interval | | Statistical significance [b] | | | |
|---|---|---|---|---|---|---|
| | | | Lognormal [c] | | Normal [d] | |
| | T1 (MYA) [a] | T2 (MYA) [a] | T2>T1 | T1>T2 | T2>T1 | T1>T2 |

*(b) Data for isoforms of three proteins*

| | | | | | | |
|---|---|---|---|---|---|---|
| Alpha actinin 1 | 138–261–495 | 62–117–222 | | | | |
| Alpha actinin 2 | 43–84–163 | 186–363–709 | * | | | |
| Alpha actinin 4 | 141–210–312 | 105–156–232 | | | | |
| Lactate dehydrogenase A | 63–114–206 | 151–273–494 | | | | |
| Lactate dehydrogenase B | 25–57–133 | 212–494–1147 | * | | | |
| Na-K ATPase beta 1 | 29–50–88 | 361–632–1107 | * | | * | |
| Na-K ATPase beta 2 | 7–14–27 | 1106–2190–4337 | * | | * | |
| Na-K ATPase beta 3 | 122–173–244 | 135–192–271 | | | | |

[a] T1 is the primate–rodent divergence and T2 is the bird–mammal divergence.
[b] Statistical significance is based on whether the 95% confidence intervals for the two times overlap (= not significant, left blank) or not (= significant, shown with an asterisk).
[c] The lognormal test results are based on the geometric means and confidence intervals.
[d] The normal test results are based on the arithmetic means and confidence intervals (which are not shown).